# BUILDING RESILIENT INFORMATION SYSTEMS FOR CHILD NUTRITION IN POST-CONFLICT SRI LANKA DURING COVID-19 PANDEMIC

Pamod Amarakoon, University of Colombo, Sri Lanka, pamodm@gmail.com

Jørn Braa, University of Oslo, Norway

Sundeep Sahay, University of Oslo, Norway

Lakmini Magodarathna, Ministry of Health, Colombo, Sri Lanka

Rajeev Moorthy, Ministry of Health, Colombo, Sri Lanka

**Abstract:** Post-conflict, low-resource settings are menaced with challenges related to low-resources, economic and social instability. The objective of the study is to understand the socio-technical determinants of resilience of resilience of routine information systems a backdrop of an implementation of a mobile-based nutrition information system in a post-conflict district in Sri Lanka. The longitudinal events in the study spans across several years into the period of COVID-19 pandemic and tries to understand the process of developing resilience of in a vulnerable district. The qualitative study deploys interviews, observations and document analysis for collection of empirical data. The case study reveals the long-standing capacity building, leadership and local governance, multisector collaboration, platform resilience and empowering of field health staff contribute in building resilience in everyday context. The empirical insights include the mechanisms in building resilience in routine system in low resource settings while promoting data quality and data use at field level.

## 1. INTRODUCTION

Low and middle-income countries (LMICs) are likely exposed to some degree of local or international conflicts at certain period during the last century. Wars or conflicts has a negative impact on country's infrastructure which severely hampers the development of the country (Hoeffler, 1999). The low resource setting observed in post-conflict LMICs paralyses the health system which needs significant level of effort for revival. Childhood malnutrition is a generally a chronic condition observed due to failure of health and other sectors in a low-resource post-conflict context ("Study on the Impact of Armed Conflicts on the Nutritional Situation of Children," n.d.). Some countries have attempted coordinated targeted interventions to malnourished children with multisector collaboration which has proven to be successful("Multi-Sector Convergence Approach to Reducing Malnutrition in Lao PDR," n.d.).

Preventive healthcare service is a major constituent of a health system of a country where provision of care is mainly community-based and provided through field health workers. Field health staff is conventionally assigned the task of gathering data of delivery of services and status of healthcare. However, the data collected and shared by field health staff is traditionally transmitted upward along the health administrative hierarchy without much supervision which can challenge the quality of the data on which public health indicators are based on (Mitsunaga et al., 2013). Another alarming situation in most of the low and middle-income countries is the lack of use of collected data at field level to enhance the healthcare delivery (Frontline Health Workers Coalition, 2016).

Digital health solutions are considered to be effective in providing transparency of data transmission and making data available to the stakeholders in a fast and efficient manner. Use of mobile technology in particular, has immensely contributed for real time data transfer between field level and healthcare managers and been increasingly implemented in low and middle-income country contexts (Tegegne et al., 2018). Cross-sector collaboration is increasingly observed across many fields and use of digital technologies has been a growing concern to make the process efficient (Lu, Zhang, & Meng, 2010). Implementation of mobile solutions for field health services has been generally challenging in the context of low- and middle-income countries that has raised increased concerns on careful planning and proper use of resources (Dharmayat et al., 2019). In addition, implementation of health information systems requires the systems to adapt to changes in the sociotechnical environment for it to operate smoothly (Yen, McAlearney, Sieck, Hefner, & Huerta, 2017). Establishment of such resilience in a vulnerable setting involves complex interplay between stakeholders and sociotechnical environment.

The study is motivated by a recent attempt of implementing a mobile based information system for nutrition monitoring at community level in Sri Lanka. The system was targeted at enhancing quality of data collected by field health staff and to enhance the data use at community level. The system was implemented in a low-resource post-conflict district of Jaffna in Northern Sri Lanka which was considered as a challenging context to establish an information system for routine use. The COVID-19 pandemic that struck the country in 2020 caused a major challenge to routine information flow in general. The study focusses at exploring deep into the broader research question of identifying socio-technical determinants of resilience of routine health information systems in a post-conflict setting. This is done through a longitudinal study of events took place in the implementation of the mobile-based information systems over a period of 3 years.

## 2. CONCEPTUAL FRAMING: VULNERABILITY & RESILIENCE

In this paper we try to focus on analyzing the factors involved in establishment of a m-Health solution in a post-conflict setting of a low and middle-income country in a manner in which it could sustain against existing and the new adverse conditions in which it is implemented. To achieve this, we contemplate on vulnerability and resilience to form the theoretical foundation.

### 2.1. Vulnerability

Vulnerability is defined as an internal risk factor of the subject or a system that is exposed to a hazard and corresponds to its intrinsic tendency to be affected, or susceptible to damage. It represents the physical, economic, social susceptibility or tendency of a community to damage in the case a threatening circumstance of natural or anthropogenic origin (Cardona, 2003; Emrich & Cutter, 2011). Vulnerability is a term which could not be confined to a specific domain or for specific group of people. It is a concept which could affect an individual or a group of people and can span across multiple domains from sociology, economics, geography, health and nutrition (Alwang, Siegel, & Jorgensen, 2001). There could be a central exposure which could make a context or society vulnerable. The degree and duration of vulnerability depends on the exposure (Kasperson, Kasperson, Turner, Hsieh, & Schiller, 2005). It is important to note that while sensitivity of a community or context to the ill-effects of an exposure suggest vulnerability, Adger defines social resilience as the ability to cope the stressors caused by changes in socio-political environment (Adger, 2000).

### 2.2. Resilience

Resilience could be broadly defined as a capacity of a system to cope with changes in its external environment (Heeks & Ospina, 2019a). In information system domain, resilience has been largely highlighted as a system property (Zhang & Lin, 2010). In certain instances resilience could be measured objectively by negative impact of an external stressor on system performance combined with time to recover to normal performance (Zobel & Khansa, 2012).Health system resilience is defined as capacity of health actors, institutions, and populations to prepare for and effectively respond to crises; maintain core functions when a crisis hits; and, informed by lessons learned during the crisis, reorganize if conditions require it (Kruk, Myers, Varpilah, & Dahn, 2015). World Health Organization defines health information systems as one of the building blocks of a health system. Therefore, it is fair to assume that the definition applied for resilience of broader health system also impacts on the domain of health information systems ("WHO | Monitoring the building blocks of health systems: a handbook of indicators and their measurement strategies," 2014). Heeks et. al highlights robustness, self-organizing and learning as foundational attributes of a resilient system while enabling attributes such as redundancy, rapidity, scale, equality, diversity and flexibility could support the establishment of resilience.

A health information system requires resilience of the health system as well as the resilience of the information system to function. Gilson et al highlights the role of routine governance structures in establishing everyday resilience of a district health system in African context. But highlights the necessity of leadership which is sensitive to district context in sustaining the resilience(Gilson et al., 2017). In the context of information systems, resilience has mainly been highlighted as a system property (Hollnagel, Woods, & Leveson, 2006). Heeks et al. identified the lack of operationalised conceptualization for resilience in the IS literature and develops a framework of foundational and enabling attribiutes with a set of markers for conceptualizing the resilience in a given context (Heeks & Ospina, 2019b).

While most of these studies focuses on general use of ICT in a community of an organizational context there is gaps in literature on how resilience could be established in routine information systems to promote data use in field health workers. Our study aims at identifying the gap in IS literature on this specific context which is essential for sustaining the health information flow of a country.

## 3. METHODS

We adopted case study methodology to explore the context prevailing within a district in Sri Lanka in the context of implementation of a mobile based nutrition information system. As highlighted by Yin, case study method facilitates empirical enquiry into the real-life scenario that happens with involvement of stakeholders at district, subdistrict and field health worker levels (Yin, 2014).

All authors of this paper were involved with the research context in multiple dimensions. One of the authors (PI) is a specialist in health informatics who was the lead health informatician in designing and implementing the mobile based nutrition information system in 4 districts of Sri Lanka. One of the authors was the Medical Officer of Maternal and Child Health (MO-MCH) of the Jaffna district who served at the managerial capacity of public health at district level. Another author is the Director of Nutrition Division of Ministry of Health (MoH) who provided oversight at national level. Two other authors are senior professors from University of Oslo, Norway who have closely been in contact with training specialists in health informatics in Sri Lanka and also was providing guidance in implementation of the project as well as physically attended field visits for qualitative research studies in Jaffna district.

As proposed by Yin, we devised multiple qualitative methods to explore the research questions of the study. As the first method we documented the lived experience of the implementers on the design and implementation in the Jaffna district which were included in the form of narratives for the study. Next qualitative method devised in the studies was in-depth interviews. These were conducted with 3 health informatics and public health experts at national level involved with implementation of the nutrition system, 3 district level administrators and public health experts, 1 provincial level public health expert of the Northern province, 2 members of Jaffna district core team implementing the system. Observations were made during the initial discussions on design and implementation within the Jaffna district, training programs conducted for end users, monthly meetings held at Medical Officer of Health (MOH) offices, a field level review conducted by national team and field visits done at field health clinics where the data capture took place. As the fourth qualitative data collection method documents and meeting notes of monthly meetings conducted in the district were analysed. In addition, 3 focus group discussions were conducted with the public health midwives who were the end-users of the system. The data collected from the above qualitative methods were documented and structured to formulate the case study presented in the next section.

## 4. CASE STUDY

### 4.1. Background

Sri Lanka is a country with a well-established public health system. Public health indicators of the country have outperformed many countries in the region and are comparable to levels of developed countries (World Health Organization, 2017). In spite of substantial performance in many aspects of preventive healthcare sector, indicators related to nutrition have failed to stay in par with others. Sri Lanka has significant percentages of malnourished children which has resulted in grave consequences to country's development over the last few decades (Jayatissa, 2011). It has been well understood that one major cause of lack of proper intervention to malnutrition was the disconnect among is multisector stakeholders operating at field level to provide efficient interventions.

The presidential secretariat of Sri Lanka launched the multisector action plan for nutrition to provide targeted intervention for families of malnourished children with involvement of all sectors related to social services provision in addition to the health domain. The objective was to establish closer linkages between multisector stakeholders at district and subdistrict level to provide targeted nutrition interventions. It was also decided to design and implement a field level mobile nutrition data collection tool to quality of data collection and data use at field level. The system was designed on free and open-source platform DHIS2 as the backend solution which communicates with a custom-made mobile application. The information system was deployed in three districts in Sri Lanka during the pilot phase from year 2016 to 2018.

Jaffna district located in the northern province of Sri Lanka is an area heavily menaced by the conflicts that prevailed over three decades ("Sri Lank's northern province poorer, undeveloped after 26-year civil war with Tamil Tigers," n.d.). The district, which is now in the post-conflict era was still having much issues related to infrastructure and technology in addition to a population living with memories of conflict era.

### 4.2. Capacity Building in Health Informatics

In the Ministry of Health Sri Lanka, numerous initiatives in administration and public health are driven by medical doctors specializing in the respected fields. Masters and doctorate programs in Community Medicine and Medical Administration are classic examples for this path. University of Colombo, Sri Lanka designed the Postgraduate Master's program in Biomedical Informatics with the support of University of Oslo in 2009 to train medically qualified doctors in health informatics who were to be absorbed in to ministry of health operations once they were qualified to serve at national and regional level for expansion of information systems. Over the last decade, they have been serving at various levels in the ministry of health hierarchy supporting implementation of information systems and well as for building capacity. In fact, the design and implementation of mobile-based nutrition monitoring system was pioneered by a graduate of the master's program.

The country also mastered in using generic platforms for designing health information systems over the years. The free and open-source health management information system platform DHIS2 is a classic example which is now being utilized by over 10 programs in the ministry of health at national level to streamline the health information management. The country has developed capacity on this platform at all levels to sustain major implementations and has kept close contacts with the DHIS2 core team at University of Oslo, Norway to ensure sustainability of support and deployment. The implementations ranges from standalone DHIS2 platform based systems to wide variety of integrations with existing solutions to establish an ecosystem of digital solutions in the country.

### 4.3. Digital Intervention

The medical officer of maternal and child health (MO-MCH) is the officer in charge at district level to implement and monitor maternal and child health activities including nutrition at district level. The principal investigator and his team in MoH designed a mobile based nutrition monitoring program based on DHIS2 as the platform. In late 2017 the MO-MCH of Jaffna district showed interest in implementing the system in the district to utilize mobile devices received for field health staff for another project. He raised the importance of strengthening multisector collaboration through the IS. In the initial discussions probable challenges for the implementation such as building capacity, supervision, infrastructure, resources for provision of support were identified as main challenges in post-conflict Jaffna districts. The MO-MCH discussed with the provincial and district health administrators, administrators of other sectors outside of the MoH at district level and the development partners such as UNICEF and World Food Program to obtain support and resources for the implementation within a short span of time. A major focus of the nutrition information system was to establish multi-sector collaboration of stakeholders operating at district level for addressing malnutrition. Therefore, he organized rounds of discussions with other government and non-government stakeholders to obtain the support. Tamil was the main language used in the Jaffna district which was posing a challenge to obtain trainers from National level who were mostly not fluent in Tamil. Therefore, he organized a training program for district core team with the help of the central implementation team to build training capacity at the district level. He and his team were able to conduct district level training programs to train the public health midwives (PHMs) on how to use the mobile data collection tool for nutrition within a period of two months followed by which the system was implemented in the entire district to collect data at field level. The team was also able to advocate the health administrators in the district and provinces to communicate the multisector collaborations on nutrition with the non-health sector government officials. The mobile application supported field health workers on their routine practice by providing snapshot idea of nutritional status of children under her care (Figure 1). The immediate supervisors of field health staff, the district and national level health managers could monitor the performance of the nutrition program using the system dashboards (Figure 2).

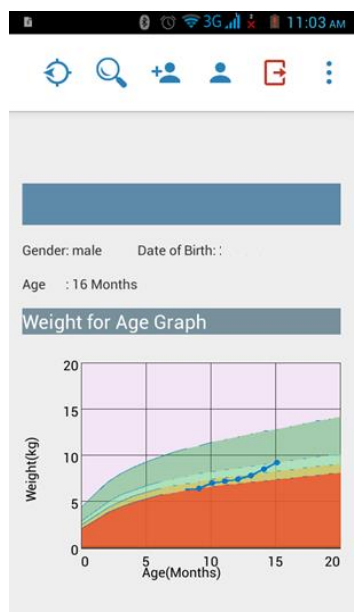

Figure 1: Weight for Age graph of a child which provides snapshot idea of progress of weight in mobile app

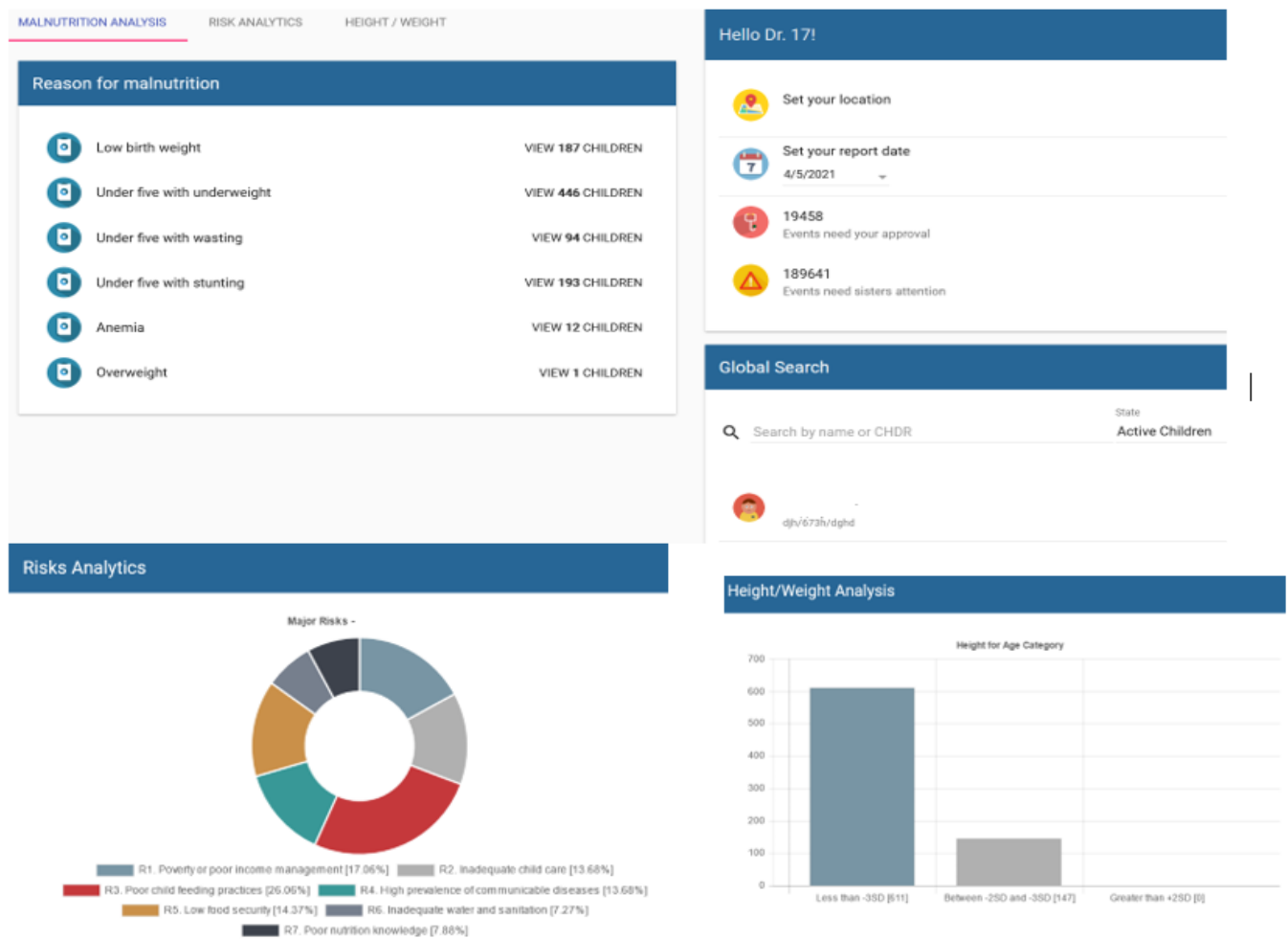

Figure 2: Dashboards of custom web application available to health managers

### 4.4. Post-implementation Phase

The lack of human resources for field health staff resulted in high turnover of health staff in low resource districts of Jaffna which necessitated conducting refresher trainings. The MO-MCH devised a formal training program at district level for newly joined field health staff prior to onboarding in field level services. The distance from the capital city and the language barrier were major challenges for conducting training and support for new users from central level. This led to significant frustration amongst end-users in the post-implementation phase. The district level team was sensitive to this growing issue and came up with a digital solution which in turn reused the mobile device provided to them. The team created a group chatroom using the instant messaging platform "Viber" for troubleshooting of end users who were requested to post their issues in Tamil language which were promptly answered by district team. They also included two members from the national team for advanced support if required. The use of instant messaging platform in turn provided peer education to the fellow end-users by learning from the issues encountered by others.

Couple of months following the initial implementation, the district team noted that the number of events of malnutrition which were recorded in the system were not up to the expected levels and also that there was considerable amount of data quality-related issues. The team analyzed the issue and attributed this development to the lack of supervision between the district and field level. To address this issue, the district team proposed to include discussion on findings from the system as an agenda item in the monthly supervisory conferences the medical officers in public health were conducting with the field health workers. Following this approach, the subdistrict team led by the medical officer of public health discussed the number of nutrition events captured as well as the quality of the data captured by the field health workers which essentially provided a feedback of data capture through the system. The field staff was also encouraged to use the dashboards in the monthly conferences which were generated based on the data collected at field level (Figure 3). This provided an instant feedback to the collected data as well as enhanced the culture of data use at field level.

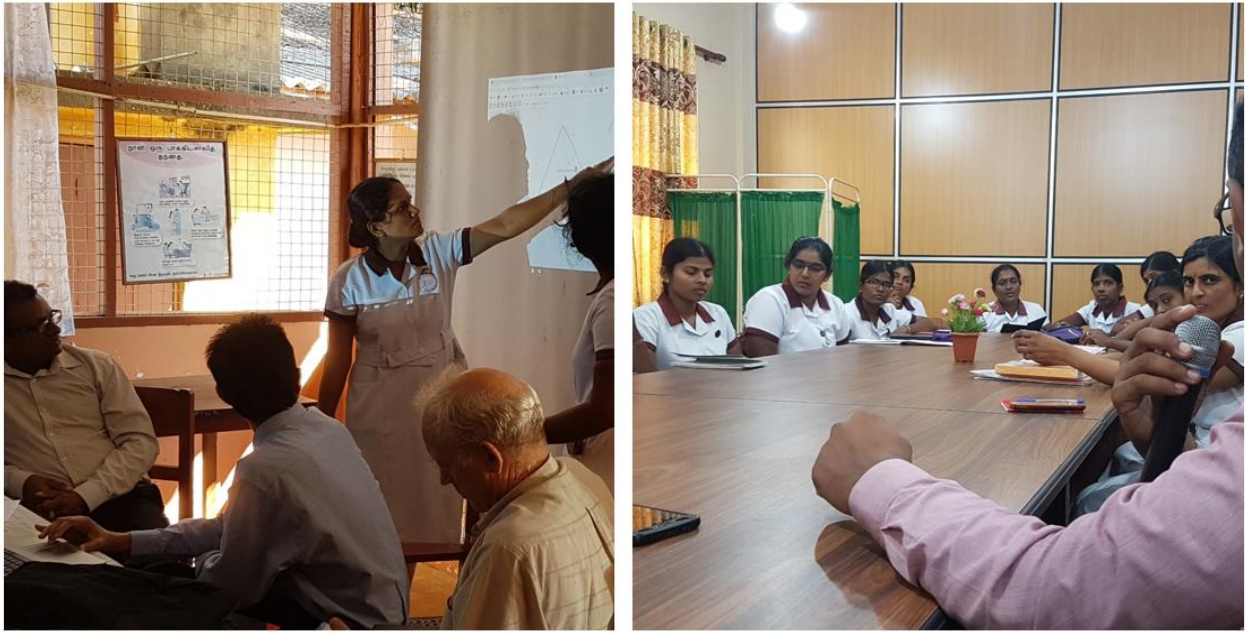

Figure 3: Monthly conferences happening at subdistrict level

The District team took one step forward in assessing the use of the information system by obtaining an unbiased objective review by inviting a team of health informaticians from national level to conduct a review of the implementation. The national team included several medical officers in health informatics as well as an international expert in public health information system implementation from University of Oslo, Norway. Their supervisory visit included visits to district health offices, field weighing clinics and the regional director of health services office to observe the practices of data collection and use and provided their feedback the staff at all levels.

### 4.5. Major Setbacks & Strategies Adopted

About two years following the initial implementation in the Jaffna district, there were complaints of non-functioning mobile devices reported from several areas. The issue led to not receiving data at all from several areas in the district. The reason was attributed to non-functioning devices following expiry of warranty period. In spite of attempting several modes of financial sources to replace the devices, Jaffna team could not obtain the sufficient funding for this purpose. This MO-MCH organized a round of discussions with public health doctors and the end-users of the subdistricts and advocated them on the value addition to the information flow by use of the nutrition system. They also observed that most of the end users owned smart mobile devices and proposed them to use personal devices for data collection. Significant number of end users accepted this suggestion and started collecting data from their personal devices. This ensured sustainability of the information system despite malfunctioning devices. The strategy was also promoted in other districts which was highly admired at national level.

### 4.6. COVID-19 Pandemic & Beyond

Sri Lanka experienced their first wave of COVID-19 outbreak in the country in March 2020. The government implemented countrywide curfew for a duration of almost 2 months. During this time only the essential services such as health care was functioning. However, this led to obstacles in conducting field weighing sessions for nutrition monitoring which affected data entry at field level. Field health staff were also utilized for COVID prevention measures which significantly overburdened limited staff. On the other hand, lack of field interventions and the extended curfew adversely affected the nutrition of children. This imposed more pressure on requirement of a digital solution to make the routine nutrition monitoring efficient to function with less over-burdened staff striving for time. In spite of not being able to conduct the routine field health work during the curfew period the district staff took prompt measures to minimize the risk of the end-users losing their practice in data collection by constantly being active in the Viber group and discussing issues related to data entry and revising the system. In addition, they made it an opportunity to conduct review and supervision of entered data in the system by conducting data review programs in the monthly conferences even during the curfew. Another interesting fact to note is that the review meeting conducted by the MoH with the Jaffna district field staff was organized through Zoom platform during the pandemic.

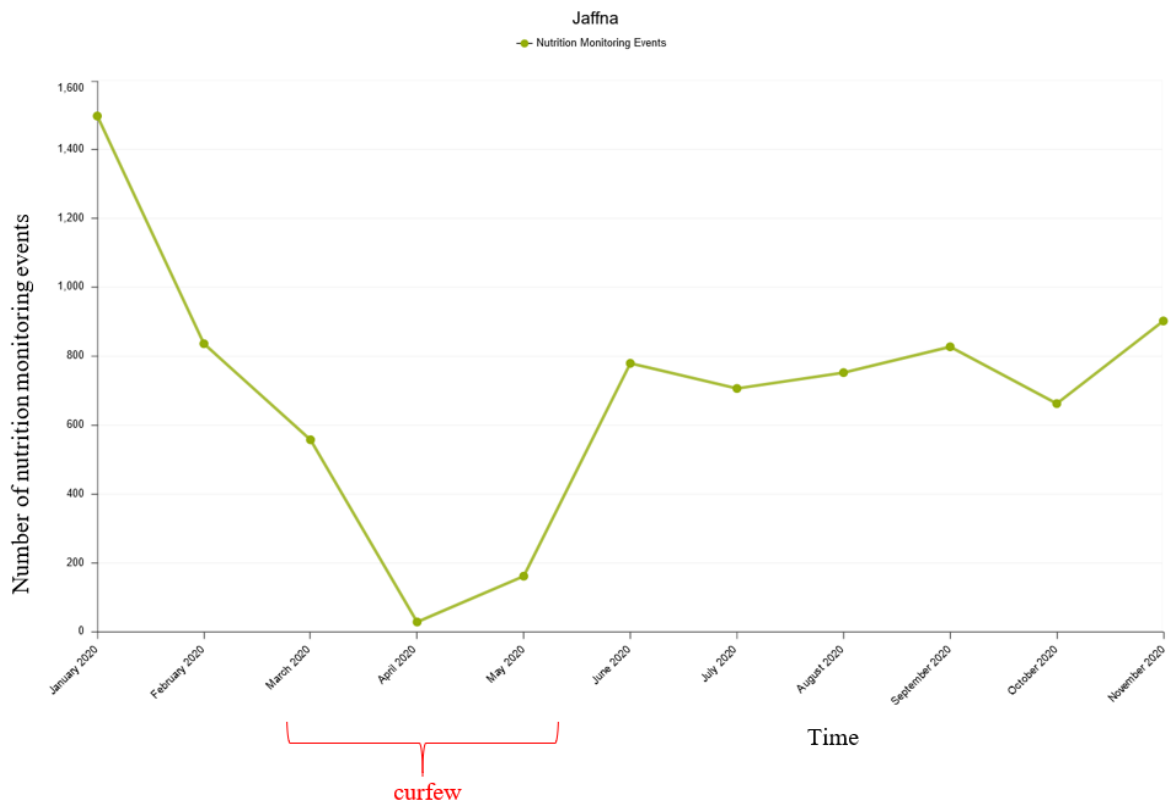

Figure 4: Graph depicting number of nutrition monitoring events captured by field health staff in Jaffna district in year 2020 (Source: District Nutrition Monitoring System, MoH – Sri Lanka)

The (Figure 4) depicts the number of nutrition monitoring events during the year 2020 in Jaffna district. It is observed that there is lack of nutrition monitoring events in first half of year 2020 which is mostly observed during the period of the countrywide curfew. However, it is interesting to note that from June 2020 onwards the events have increased and come to a plateau which continued to remain the same during the next few months amidst field health staff being occupied with COVID-19 preventive measures.

Another interesting fact to note is that the champion behind the implementation, the MO-MCH was transferred to a different post at the end of 2020. However, he handed over his duties to his successor and build capacity in the team that continued to support implementation. He continued engaging in provision of technical and user support actively and chat support channels in the next three months. His void was filled by a Medical Officer in Health Informatics graduated from the Masters programme who was oriented in the system and had the capacity to provide leadership and expertise in sustaining the system. The success behind Jaffna district team inspired the national maternal and child health program in the MoH to incorporate the remaining child health data collection fields into the mobile application to enhance the quality of child health service provision and monitoring and evaluation activities. Integration of the maternal and child health flow into the same mobile application was a major step to ease the burden duplicate data collection as well as ensuring the sustainability of the grassroot level digital data capturing process.

# 5. DISCUSSION

The paper revolves around a use case in post-conflict resource-limited district of Jaffna in Sri Lanka and how implementation of a mobile based health information system progressed amidst numerous challenges over a period of 2 years. Our analytical focus in this paper is on the relationship between the local operational mechanisms around the information system and how it contributed to the resilience of establishing a routine information system which supported enhancing data quality and use. The challenges we set out in this paper, are: "what it takes to build a resilient information system in a low-resource LMIC setting" and, within this overall case framework, we address 1) the process of building information system resilience, and 2) the bounce back-bounce forward duality of resilience, all in the context of establishing a digital IS in challenging times of COVID-19 pandemic.

The Jaffna district, in which the case study revolves around was in post-conflict stage after a 3 decade-long civil war which collapsed not only the infrastructure and livelihood, but also the general motivations and quality of life of people. Emrich and Cutter highlight that interaction of society with biophysical conditions caused by the vulnerability has significant impact on the way they adapt or build resilience in general (Emrich & Cutter, 2011). The vulnerability of economic domains led to lack of infrastructure and equipment while the vulnerability of psycho-social environment created a hostile environment for motivation and enthusiasm. Lack of capacity made the district furthermore vulnerable for embracing novelty or sustain it. This, in turn made the district vulnerable for long-term dependence for support from non-governmental organizations and development partners.

Sri Lanka envisioned about a decade ago the need of health informatics expertise and capacity to implement and sustain IS when establishing Masters program in Health Informatics. The approach of Sri Lanka on building capacity was slow and steady to build resilience against their vulnerability of capacity in implementing health information systems. The void of transfer of the champion doctor behind the implementation of IS was quickly filled by a medical doctor with expertise in health informatics which led to strengthening and continuity of building of resilience in the district.

Sri Lanka also built capacity around the DHIS2 platform steadily during the decade. The country had time to experiment with the platform and also to implement it at different scale while also developing capacity to expand the platform by designing complex, integrated mobile applications as in the case of the nutrition system. Best practices, implementation guidelines and governance mechanisms around the platform were established based on several use cases during the decade. Tiwana highlights the fact that aligning the governance with the platform's architecture improves platforms resilience (Tiwana, 2014). Therefore, the case highlights how a LMIC accumulated platform resilience at a gradual pace which produced resilience within the country to sustain advanced IS.

When analyzing the case study, we identify that there were several instances where the system was severely challenged for its survival after the initial implementation. When the quality of data as well as use of data generated from the system was questioned at district level, the justification on resources spent for establishing the IS was threatened. When the user support mechanism was not properly established there was a risk of losing the use of the system at field level. These can be highlighted as examples of stressors to the routine functioning of the information system. The mechanisms operating at the field level were able to come up with strategies to overcome these stressors to bring back the system into normal use which could be referred to as "bouncing back" to normal. This is a general property of resilient IS (Heeks & Ospina, 2019a). However, the strategies that were devised by the district team such as introduction of reviewing system performance at monthly conferences, introduction of "Viber" support group and moreover familiarizing the field staff to use an instant messaging platform for official purposes strengthened existence of nutrition system as well as contributed to creating a culture which could be reutilized in routine work. This strategy contributed to technology resilience during the times of COVID pandemic in addition to empowering end-users with a skill they can reuse for routine work. These are unique examples of systems transformation which pushed the exiting boundaries of Isa a step further than bouncing back. Müller et al. Refers to this as "bouncing forward" which could be defined as what strengthens the resilience of the system to face future stressors (Müller, Koslowski, & Accorsi, 2013).

Gilson et al. stresses the importance of everyday resilience for the sustainability of the health systems in low and middle-income countries (Gilson et al., 2017). They emphasize that having stable governance structures and adequate resources will not make the sustainability of the everyday resilience. They go on arguing that it is imperative to have new forms of leadership which empowers all levels while maintaining cordial relationship with other stakeholders to sustain the everyday resilience. Our case study highlights the initiatives undertook by the MO-MCH of the Jaffna district in introducing the information system into the low resource setting, interactions and advocacy he had with stakeholders at multiple levels across the domains and the initiatives he pioneered while empowering the district team as well as the field level staff. These steps were crucial to obtain the support of all stakeholders as well as laying a formidable foundation on which the system could establish in the district. The willingness of the maternal and child health programme of the MoH to incorporate the general MCH indicators into the existing mobile based nutrition information system provides sufficient evidence for the Ministry recognizing the prospects of this implementation to establish as a routine information system for everyday data requirements.

COVID-19 pandemic brought about major challenges to the health system as well as the frontline health staff. This is where the value of the system with everyday resilience becomes crucial. The nutrition system not only bounced back following previous stressors but bounced forward with modification such as incorporating supplementary digital solutions like instant messaging platforms. This resulted the ecosystem around the nutrition information system to further strengthen by means of conducting data quality reviews even though the field level data collection was not possible. The system empowered field health workers by having decision-making tools such growth charts at their fingertips as well as the dashboards for managerial staff. These methods brought about compliance as well as a sense of ownership to the end-users. They made the system a part of their daily work-routine which in turn consolidated the platform's everyday resilience. The reduction of nutrition monitoring events during the period of country-wide curfew was the actual reflection of service delivery. But the fact that events reaching levels of pre-covid era just after pandemic and continuing to remain that levels highlight the fact that COVID pandemic has not crippled the use of the information system. The maternal and child health reviews were conducted by the ministry in times of the pandemic using the Zoom technology which also highlights the resilience developed in use of digital platforms and the digital transformation of health system in general.

Thus, through the case study we try to explore closely the interplay between vulnerability and resilience of routine health information system through on empirical lens.

# 6.    CONCLUSION

Through this case study of implementation of a mobile based nutrition monitoring tool by field health staff, we try to capture the socio-technical determinants of information system resilience in a post-conflict LMIC setting. Long term capacity building, local governance mechanisms, platform resilience, multi-sector collaboration and empowering of end-users are crucial determinants in constructing resilience which can bounce back from routine stressors and bounce-forward to establish as a resilient routine system in the backdrop of COVID-19 pandemic.

The study contributes to the theoretical domain by understanding socio-technical determinants of resilience of routine IS and highlighting the importance of empowering field health staff for achieving digital transformation in routine system. Empirically the study provides critical insights on the process of establishing resilience in routine IS in a post-conflict low resource setting and measures of enhancing data quality and data use of at level of field health workers in such setting.